\documentstyle[12pt,psfig]{article}

\begin{document}

\title{Electron Temperatures in W51 Complex from High Resolution, Low Frequency Radio Observations }

\author{{P.K.~Srivastava$^{1}$, A.Pramesh~Rao $^{2}$}\\  
\small{$^{1}$ DAV College, Kanpur, India 208001}\thanks{pradeep@iucaa.ernet.in}\\ 
\small{$^{2}$ NCRA, TIFR, Pune, India 411007} \thanks{pramesh@ncra.tifr.res.in }} 
%\date{Accepted???.Received????;in original form ????}

\maketitle

\begin{abstract}

W51 is a giant radio complex lying along the tangent to the Sagitarius arm at a
distance of about 7kpc from Sun, with an  extension  of about  1$^\circ$  in the
sky.  It is divided into three components  A,B,C where W51A and W51B 
consist of many compact HII regions while W51C is a supernova remnant. We have made continuum radio observations
of these HII regions of the W51 complex at  240,610,1060,1400 MHz using
GMRT with lower resolution ($20''$ $\times$ $15''$) at the lowest frequency.  The observed  spectra of the prominent  thermal subcomponents  of W51
have been  fitted 
to  a free-free  emission spectrum and their physical  properties like electron
temperatures  and emission measures have been estimated. 
The electron  temperatures from  continuum  spectra 
are found to be lower than the temperatures  reported  from radio recombination
line (RRL) studies of these HII regions indicating the need for a filling factor even at this resolution. Also, the observed brightness at 240MHz is found to be higher than expected from the best fits suggesting the need for a multicomponent model for the region.

\end{abstract}


%\end{keywords}

\section{Introduction}

W51 is a giant radio emitting complex, lying along the tangent to the Sagitarius  arm at
a  distance  of about  7kpc and  having an angular size of about  1$^\circ$.
The complex consists of many HII regions of varying morphologies and
compactness,   embedded  in  an extended  and  diffuse  emission.
According to convention  adopted by Kundu \& Velusamy(1967)  the source is 
divided into 3 components, W51 A and B being a collection of compact HII
regions  while W51C is a nonthermal source that is identified with a supernova
remnant (Copetti \& Schmidt  1991).  The entire
complex  is  optically  obscured but  has  been  extensively observed  at  radio  frequencies
and the early  observations  have been reviewed by  Bieging(1975).  The
brightest  HII  regions are  found  in  W51A,  which  itself   consists  of  
two
subcomponents  G49.5-0.4 and  G49.4-0.3.  \ G49.5-0.4 is the most  luminous star
forming region in the entire  complex and has been studied in detail.  High
  resolution radio continuum  observations  (Martin 1972; Mehringer 1994;
Subrahmanyan \& Goss 1995) have identified atleast 8 prominent  compact sources in
G49.5-0.4  which are  referred to as  G49.5-0.4 { \it  a,b,...,h  } in order of
increasing right ascension.  Similarly,  G49.4-0.3  consists of atleast 3 compact
sources  G49.4-0.3 {\it a,b,c} (Martin  1972).

Radio recombination line (RRL) studies  of  G49.5-0.4  have been  used to study 
the kinematics and local thermodynamic equilibrium(LTE) electron 
temperature($T_{e}$) of this region.  H109$\alpha$ RRL observations by Wilson et
al(1970) with a resolution of $4'$ and by Pankonin et al  
(1979) with a  resolution of 
$2.6'$ show  that  $T_{e}\sim $6000K  in  G49.5-0.4.  Lower  frequency
observations for H137$\beta$ and  H166$\alpha$  lines show higher  temperatures of 8500K and 7100K respectively (Pankonin et al 1979).  Since lower
frequency lines are expected to arise from outer low density parts of the HII
regions,  this  results  suggest  that  there is a  temperature  gradient that
increases radially  outwards  (Churchwell  et al 1978).\ From  observations  of
H92$\alpha$  lines  Mehringer(1994)  concludes that $T_{e}$  varies from 4700 to
11000K  for the different components with an average  of 7800  $\pm$  1200K.  All the RRL  results quoted above have been
derived under the LTE approximation.

  The electron temperature is one of the most important parameters in understanding the
physical   properties   of  thermal  HII   regions.  Low   frequency   continuum
observations  in the optically thick regime offer a direct  estimate of the electron  temperatures.
Meter-wavelength  observations  at 410MHz by  Shaver(1969) and  at  151MHz  by   Copetti  \&
Schmidt(1991)  show that $T_{e}$ in G49.5-0.4 is
3500$\pm$700K  and  4650$\pm$500K  respectively.  Since low frequency  continuum
observations  also  arise  from the  outer  parts of HII  region,  these low temperatures 
suggest a radial  decrease  in  temperature  from the  core, which is
opposite to the trend seen in RRL  studies.  However,  Subrahmanyan  \&  Goss(1995),
from 330MHz continuum  observation with $\sim$$1'$ resolution, infer a value
of at least 7500K towards  G49.5-0.4{\it  e}, the brightest source in G49.5-0.4.
It  is possible that  the low  temperatures   obtained  by Shaver(1969) and Copetti  \&
Schmidt(1991)  is due to the poor  resolution ($>3'$) 
of their images.  

In order to resolve this discrepancy and make an improved determination of the electron  temperatures and other physical parameters in
the  component  W51A, we have  made high  resolution
continuum images of the HII regions in W51A using Giant Meterwave Radio Telescope (GMRT) at  240,610,1060,  and 1400MHz.  This is the first time that W51 has been
observed at meter-wavelengths with arcsec resolutions.

\begin{table*}
   
\caption{Observational Parameters}\vspace*{5mm}
   
\begin{tabular}{ccccc}
\hline
 Central & Date &   Bandwidth  & Best   & Primary   \\
Frequency(MHz)&  &(MHz) &Resolution  &Beam  \\
\hline
240 & 9Mar2001  & 6.25 & $19''\times14''$ @50 $^\circ$ & $114'$  \\

616 & 30Aug2001  & 12.625 & $6''\times 6''$ & $43'$ \\

1057 & 5Jun2001  & 12.625 & $3.7''\times3.2''$ @82$^\circ$ & $32'$ \\

1407 & 4Jun2001  & 13.25 & $3''\times3.5''$ @ -79$^\circ$ &  $24'$ \\ \hline

\end{tabular} \label{tab1} \end{table*}

\begin{figure} 
\begin{center}  
\psfig{file=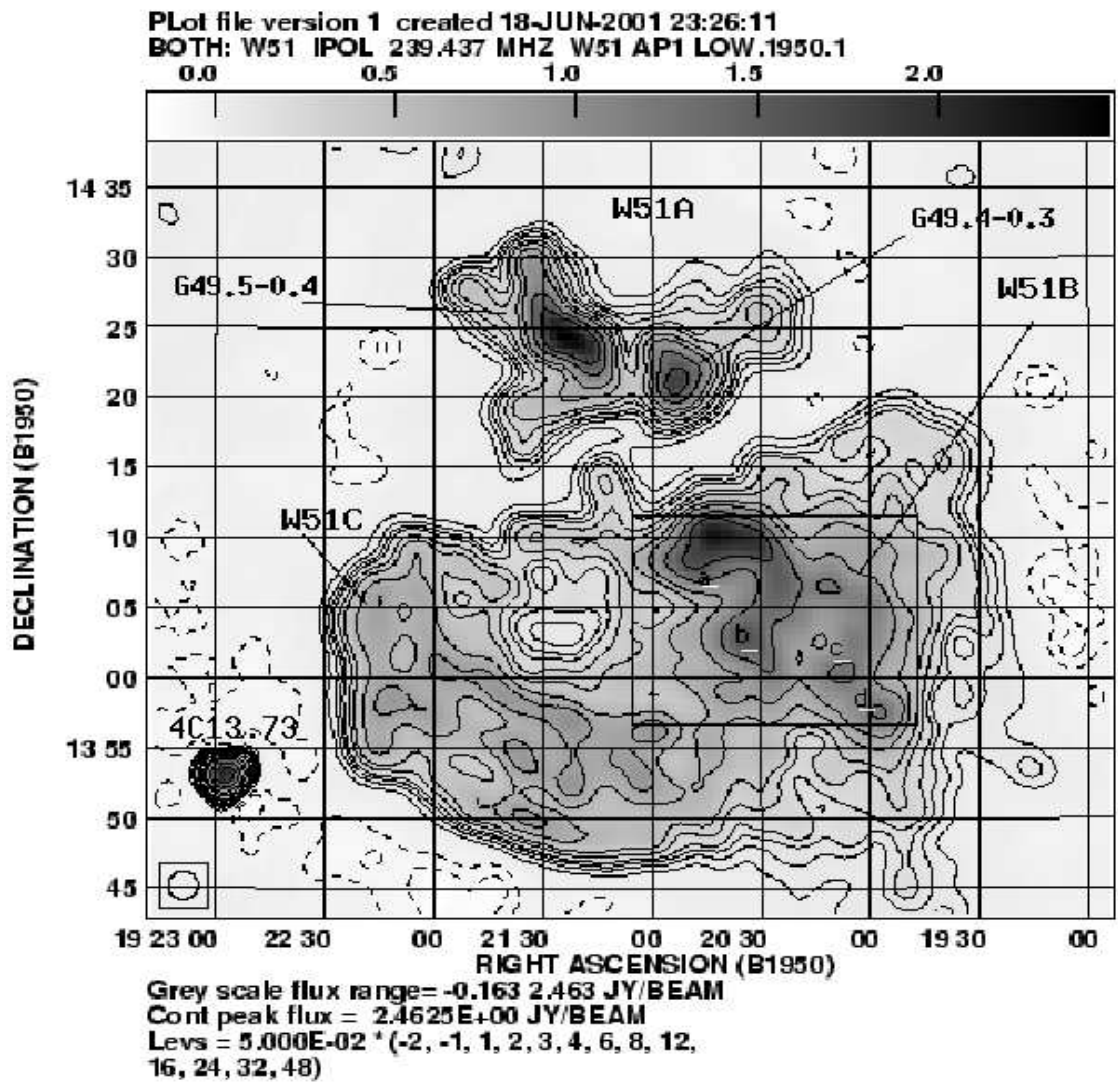,height=15cm,width=15cm}
\caption{  Map of full W51  complex  at  240MHz  with  resolution  of $2'$. The rms noise is 21mJy.}
\end{center}\label{Fig.1} 
\end{figure}

\section {GMRT Observations}

\subsection{Observations and Data Reduction}
 
Full synthesis observations of the W51 complex were made with the GMRT during
the year 2001.  Details of the  observational  parameters  are given in Table 1.
All the  observations  were  centred on W51A.  While at the higher
frequency  the fields of view were  restricted  to only  W51A,
the 240MHz  data had a field of view large  enough to cover the  entire W51
complex.  The observations were made in the standard  spectral line mode of GMRT
(128 channels, each of 128KHz  width) for nearly 8 hours at each frequency.  A band pass  calibrator,  which was also
the flux density  calibrator  (either  3C48 or 3C286 or both), was  observed  
for 15-20  minutes at the  beginning or at the end of the  observation.  The rest 
of the  observation  was devoted to W51, with a secondary calibrator  observed  
every half an hour or so. 

 In the offline  analysis, which was done using
standard  AIPS, the data were  edited, the antenna  bandpass  estimated  and the
spectral  channels  collapsed to a continuum  channel.  For
the 240MHz data, where  bandwidth  smearing is an issue, the  spectral  data was
averaged to 5 channels  of 1.25MHz  each.  The  channel  averaged  data were 
calibrated using the secondary  calibrator.  At all observing frequencies, uniformly 
weighted maps were made with the full resolution of the GMRT, cleaned,  and
self-calibrated  (two rounds of  phase and one of amplitude) to make the 
final map. For images at 240MHz, wide field  corrections  were  incorporated  
using the multiple facet imaging  option in IMAGR.  Since the resolution of GMRT
varies with frequency,
in order to compare the maps at the different frequencies, additional maps were made at the higher frequencies
with lower resolution ($20''$ $\times$ $15''$) comparable with the 
resolution of the GMRT at 240MHz and the 5 and 1.4GHz maps presented by 
Mehringer(1994). It must be noted that the {\it u-v} coverages are different in the 
different maps which could lead to uncertainties in comparing the maps at different 
frequencies. A low resolution map $2'$ of the full W51 complex was also made at 
240MHz to compare with the existing images.

During the period of these observations,  
real time measurement of the system temperature was not  
available at the GMRT.  The change in the system temperature  between the primary  calibrator 
and W51 was measured  later and the flux densities in the final maps were  
corrected for this (correction  factor ranged from 1.2 to 1.6).  All the images 
were  corrected  for the primary beam of the GMRT antennas.

\subsection{240MHz Image of the W51 Complex}

       Figure 1 shows full field map of W51 at 240MHz made with a
resolution  of $2'$ .  The overall  features agree very well with
the VLA map at 330MHz with similar resolution (Subrahmanyan  \&
Goss 1995).  The thermal sources  W51A in the north and W51B in the
centre,  and the nonthermal source  W51C are  marked in the  figure.
The  extended nonthermal emission from the supernova  remnant W51C is
clearly seen. The overall angular extent of the source is  about $45'$
and since the shortest spacing in our observations is 50 wavelengths at
240MHz, we believe that our 240MHz images are not seriously affected
by the absence of measurements at short spacings.  We have tried
to alleviate this effect by using the \lq zero spacing\rq flux option in
IMAGR and have carefully compared our integrated flux densities of
the various components with those estimated by others  ( Copetti \&
Schmidt  1991, Subrahmanyan \& Goss 1995) to convince ourselves that
our flux density estimates are correct.  Inspite of this there is some
evidence in Figure 1 for a negative bowl around the complex at about
50 mJy/beam and this has been included in the error estimate. While
this could cause us to underestimate the flux density in the extended
emission, its effect on the high resolution maps is negligible since
the beam areas are smaller. Due to various systematic uncertainities,
including the correction for $T_{sys}$, we believe that our flux density
estimates are accurate to about $10\%$. This was further verified using
the source 4C13.73 at the south-east boundary of W51. The source which
is resolved into a  double in the high resolution 240MHz images,  has a
flux density of 4.2$\pm$0.4 Jy at 240MHz which agrees within errors with
estimates based on previous measurements at 151 and 330MHz (Copetti \&
Schmidt 1991, Subrahmanyan(private communication)).

 The flux density observed for the W51 complex at 240MHz (Figure 1) is 305$\pm$ 30 Jy, 
out of which 46$\pm$ 4Jy lies in W51A.  The  integrated  flux  densities  in the 
two subcomponents  of  W51A, G49.5-0.4 and G49.4-0.3, are 29$\pm$3 and 17$\pm$2Jy 
respectively.  These flux densities are consistent with 330MHz measurements 
of Subrahmanyan \& Goss  (1995)  with the proviso  that  these are  
thermal  sources  that become optically  thick below  1GHz.  
The  separation  of the rest of W51 into W51B and
W51C is  slightly  subjective  and if done the way as shown in Figure  1, the total
flux  densities  in W51B and  W51C  are  110$\pm$15 and 150$\pm$ 20 Jy 
respectively. The regions  (marked  {\it  a,b,c,d})  in W51B  corresponding  to 
the  sources G49.2-0.4,  G49.1-0.4 and G48.9-0.3 ({\it c,d}) respectively, along 
with the connecting bridge of  emission  seen at  higher  frequencies  
(Subrahmanyan  \& Goss  1995, Altenhoff  et al 1978), are visible in our map 
too, though all the  condensations are fainter than in the 330MHz map of 
Subrahmanyan \& Goss, confirming that they are thermal  sources.
 
 Our estimate of the  integrated  flux density of W51C is higher than that  
of Subrahmanyan \& Goss(1995) but this could be  due to different
regions being included in W51C.  A more useful  quantity is the average  surface
brightness  over W51C at 240MHz which, in the units used by Subrahmanyan \& Goss (1995),
we  estimate  to be 0.22 Jy/beam of $1'$ FWHM, if we  include  the entire
extent of the source including the fainter  extensions to the south west.  If we
exclude this region, since this may be more  appropriate  for comparing with the
lower  dynamic  range  single  dish  maps and the 151MHz  maps of  Copetti  and
Schmidt(1991),  we find that average surface  brightness is 0.29 Jy/beam of $1'$ FWHM. These
values lie between estimated values at 151 MHz (Copetti \& Schmidt 1991) and 330
MHz  (Subrahmanyan \& Goss 1995) supporting their hypothesis that the brightness
temperature  of W51C starts to fall below 410 MHz,  probably  due to internal or
foreground  free-free  absorption.  

\subsection{Matched Resolution Images at Other Frequencies}

	Using the GMRT data at 240MHz, the highest resolution that could be achieved on
W51 was $20''$ $\times$ $15''$ ($0.66pc \times 0.5pc$ at a distance of 7kpc) at position angle 45$^\circ$.  A number of compact sources in and around W51 are
seen in the high resolution map and 4C13.73 is clearly resolved as a double source.
While much of the extended emission
of W51B and C is resolved out,  W51A, being more compact, is
seen clearly and can be used for quantitative analysis (Figure 2d). While the higher
frequency GMRT data can support higher angular resolution, maps with resolution comparable  
to the maximum resolution of 240MHz maps were
made from the  610,1060  and 1400 MHz  data and are shown in Figure 2.
The  features  seen  in our 1400 and 1060 MHz  maps  agree  well  with  the 1400MHz  VLA map
of comparable resolution (Mehringer 1994).  The  regions  G49.5-0.4{\it  a-h} identified by
Mehringer are also seen in our maps and all of them are thermal  sources.  
Comparing the flux densities of individual sources, the GMRT flux densities at 1400 MHz seem
to be about $20\% $ lower than estimated by Mehringer. It is not clear if this is due
to the different {\it u-v} coverages, slight difference in centre frequencies, or the different regions being integrated to get the flux densities, but we 
have absorbed this in the error estimates. All the individual 
components  are also seen in the 610MHz  map but they  appear  fainter,
clearly  showing  self  absorption.  In the 240MHz map,  detectable
emission is present at the location of many of these  HII regions  though it is not 
obvious  that the  emission is from the same  physical  region as that at higher 
frequencies. The 240MHz map shows diffuse extended emission not seen in the other 
maps which is probably due to nonthermal galactic emission which is detected due to 
the relatively better short spacing coverage at 240MHz. 

We have used the $20''$$\times$$15''$ images for quantitative analysis.
The peak surface brightness (in units of mJy/beam) and the integrated  flux density of the 
HII regions at all the frequencies of observation are estimated and presented in Table 2.
The integrated flux of the components was estimated by integrating the  flux per 
beam in the maps over the extent of the component defined by appropriate boxes.
The definition of boxes defining the region of a component is a bit subjective
when it is in a complex region, but we have ensured that all the maps were 
integrated over identical regions. In Table 2 we have also given the peak surface 
brighness of the components of G49.5-0.4 at 4680 MHz using VLA  maps of Mehringer(1994) of comparable resolution 
(courtesy archives of the Astronomical Data Image Library, maintained by the NCSA). 

 At 240MHz, where the surface brightness of many of the components is small and 
the background is high, we have estimated the peak of the sources by taking 
slices across the regions in different directions and estimating the height of the 
source above the surrounding region. While this was simple for isolated sources, 
separating the background and the source
was often difficult for the sources in complex regions like G49.5-0.4 {\it c,d,e}. 
Such a background correction is appropriate when the surrounding flux density comes 
from regions around or in front of the source. A full justification for or against 
this procedure requires a detailed understanding of the relative location of all 
the emitting and absorbing regions, which we do not have. However, we feel that this 
subtraction is justified since there is likely to be foreground emission for the 
components in complex regions and we were concerned about possible biases in the 
240MHz map due to its better short
spacing coverage. Making maps at 240MHz with the short spacings removed did not 
help since the extended emission caused a negative bowl around the extended 
emission which still needed to be corrected. The estimation of the background and 
its subtraction produces uncertainties that are reflected in the error bars put on 
the estimated peak surface brightness (Figures 4,5).
A feature of the 240MHz map is the diffused emission to the east and the south-
east of the W51A  complex. This  emission is seen in the VLA 330MHz maps, but is 
not so prominent at the higher  frequency  maps of both the VLA and the GMRT.

While the full resolution images at 610, 1060 and 1420MHz and their analysis will be presented elsewhere, we present in Fig.3, a high resolution $3.7''$$\times$$3.2''$ image of W51A at 1060MHz. The complex nature of the region with 
the compact sources, shells and diffuse emission, seen in the  high resolution 
images at 5 and 8GHz (Mehringer, 1994), are also seen in our map though with lower 
brightness, with the compact regions showing evidence
for self absorption.

\begin{figure}    
     
\centerline{\hbox{
\psfig{figure=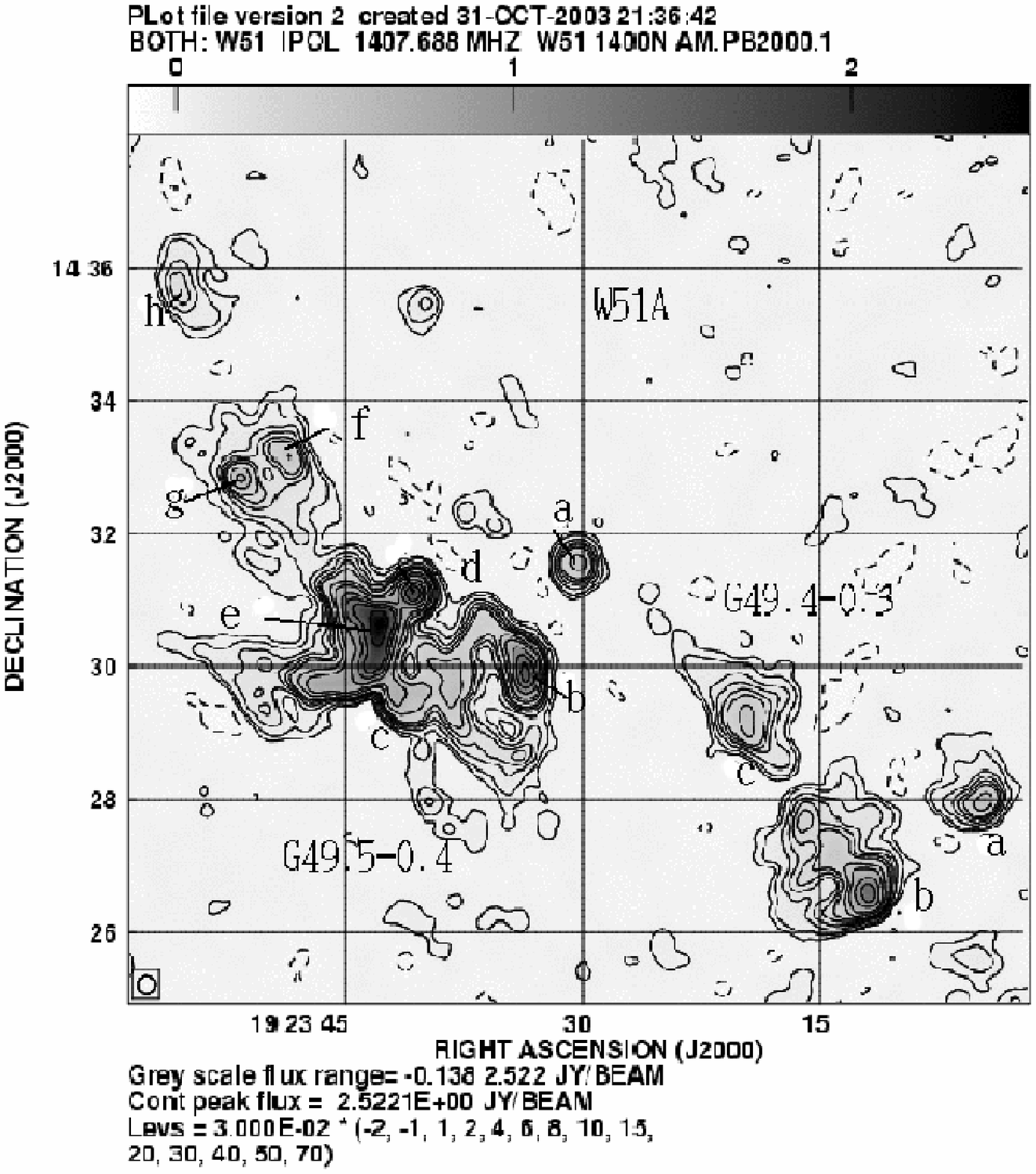,height=8cm,width=8cm}  
\psfig{figure=Fig2b.PS,height=8cm,width=8cm}}}  
\centerline{\hbox{
\psfig{figure=Fig2c.PS,height=8cm,width=8cm}  
\psfig{figure=Fig2d.PS,height=8cm,width=8cm}}}  

\caption{Maps of W51A at 1400,1060,610,240MHz made at appoximately same 
resolution of $20'' \times 15''$. The rms noise in above maps is 12,10,6,10mJy respectively }

\end{figure}

\begin{table*}
                    
\caption{Observed Peak Flux Density(mJy)/beam ($I(\nu)$) for Beam of $20''$$ \times$$ 15''$ and Integrated Flux(Jy) for HII components in W51A}  
\vspace*{5mm}

\begin{tabular}{cccccccccc} \hline   
Frequency & \multicolumn{2}{c}{240MHz} & \multicolumn{2}{c}{610MHz} & 
\multicolumn{2}{c}{1060MHz} & \multicolumn{2}{c}{1400MHz} & 4860MHz* \\

Source & $I(\nu)$ & Flux & $I(\nu)$ & Flux & $I(\nu)$ &Flux &$I(\nu)$ &Flux & $I(\nu)$        \\ \hline
G49.5-0.4a&66&0.27&160&0.46&440&0.93&590&1.29&530 \\

b &120&0.79&330&1.20&1310&3.62&2060&5.11&3040 \\

c &120&0.62&260&0.71&790&2.04&990&1.97&1130 \\

d &36&0.42&340&0.82&1430&2.81&2210&3.24&9030 \\

e &90&1.28&430&2.37&1750&7.97&3040&12.3&10700 \\

f &60&0.52&180&0.62&440&1.18&480&1.26&620 \\

g &50&0.49&200&0.55&640&1.50&740&1.42&1000 \\

h &110&0.56&120&0.41&250&0.66&240&0.66&270 \\

\hline

G49.4-0.3a & 80 & & 230 & & 490 & & 630 & & \\

b & 80 & & 310 & & 1110 & & 1630 & & \\

c & 80 & & 240 & & 500 & & 540 & & \\ \hline

\end{tabular}

\flushleft\small{ * From Mehringer(1994)}

\label{tab2} \end{table*}

\section{Measured Parameters and Physical Conditions of W51A Components}

Both the  components  of W51A,  G49.5-0.4  and  G49.4-0.3,  consists  of several
compact HII sources  surrounded  by diffused  ionised gas (Goss \& Shaver  1970,
 Mehringer  1994).  The  emission  from the entire  W51A  region is  
thermal and the sources are seen to be  optically  thin at  frequencies greater 
than 5GHz
(Martin1972,Mehringer1994).The  flux density due to continuum free-free emission
from a homogeneous  and isothermal
spherical clouds is given by (Hjellming et al 1969)

\begin{equation}  S = 3.07 \times  10^{-2}  T_{e} \ \nu^2  \Omega  (1-e^{-\tau})
 \end{equation}

 where $S$ is integrated flux density in Jy, $T_{e}$ is electron  temperature in
Kelvin, $\nu $ is frequency in MHz,  $\Omega $ is solid angle  subtended  by the
source in  steradians  and $\tau $ is optical depth along the line of sight.  At
radio  wavelengths,  optical  depth  $\tau$ can be   approximated as (Mezger \&
Henderson 1967) 

\begin{equation}  \tau \sim 1.643  \times  10^5  \nu^{-2.1}  EM \  T_{e}^{-1.35}
  \end{equation}

where $EM$ is  emission  measure  in  cm$^{-6}$pc and the frequency is in MHz.  
For  resolved  sources  with more complex structure, the brightness as measured by the flux density per beam is the more appropriate quantity and 
following  the  procedure  adopted by Wood \& Churchwell
(1989), we consider the value of brightness at the
position    of   maximum    intensity    as   the   best    representative    of
$S$/$\Omega$. In Table 2 we have listed the observed peak flux  density for
a  gaussian beam of FWHM of $20''$$\times$$15''$.

The observed peak brightness temperature is equal to the true brightness temperature only if the source is well resolved and does not have fine structure smaller than the beam. If the source is unresolved or has fine structure, the observed peak brightness temperature is less than the true brightness temperature by an unknown factor that is determined by the ratio of the effective solid angle of the source to the solid angle of the observing beam. The main aim of high resolution observations of these sources at low frequencies is to resolve the source and get a direct estimate of the temperature of the emitting region without the uncertainty of the filling factor. The VLA observations of Mehringer (1994) and our own high resolution observations at 1400MHz suggest that many of the components are resolved with a $20''\times15''$ beam and so could be candidates for a direct determination of the brightness temperature.

For a beam of $20''\times 15''$, the observed peak flux density per beam $S_{p}$/$\Omega_{20''\times15''}$, estimated from the images, written as $I_{\nu}$, can be got from Eqs 1 and 2 as

\begin{equation} I_{\nu}=S_{p}/\Omega_{20''\times15''}= a \nu^{2} (1-e^{-b\nu^{-2.1}}) \end{equation}

where $a = 2.45  \times  10^{-4}  T_{e}$ , and $b = 1.643  \times  10^{5}  EM \
T_{e}^{-1.35}$

The electron  temperature  $T_{e}$ and emission measure $EM$ were estimated
by least-square  fitting of the observed data (Table 2) into Eq.3 and the 
estimated $T_{e}$ and $EM$ are presented in Table 3. The plots of $S_{p}$/$\Omega_{20''\times15''}$ against frequency  for all the sources in G49.5-0.4  and G49.4-0.3 
are shown in Figures 4 and 5.  As is clear from the figures, in most of the
sources the data agrees well with the assumption of a thermal source that is 
optically thick at metre wavelengths and the data is good enough to get reliable
estimates of both the electron  temperature and the emission measure. Some of the sources show excess emission at 240MHz. The main exception is the source 
G49.5-0.4{\it h} (Fig 4) which does not show an unambiguous low frequency turn over
and suggests that this source is optically thin at most of the GMRT frequencies or has greatly enhanced brightness at 240MHz.

\subsection{Electron Temperature}

The electron  temperatures for all the HII
regions,   as  obtained  from  above   least-square   fitting,  range  between
2100-5600K.  These temperatures are lower than expected from standard models 
of HII regions and from RRL studies but seem to be consistent with the
literature where low frequency radio observations tend to give lower temperatures
than other techniques (Shaver 1969, Kassim et al 1989 and Coppetti et al 1991). The  low  frequency
continuum observations at 410MHz (Shaver1969) and at 151MHz (Copetti \& Schmidt 1991) 
estimated temperatures of about 4650K and 3500K respectively for the 
entire W51A complex and
it has been argued that these low temperatures are perhaps due to the low resolution
($>3'$)  of their observations.  Subrahmanyan \& Goss(1995)  from their 330MHz 
continuum observation with $1'$ resolution inferred that  
$T_{e}\sim 7800K$ for  G49.5-0.4{\it d,e} components. Our estimate of 
$5600 \pm 400$ for the temperature of G49.5-0.4{\it e}, is consistent
with their direct estimate of $6200 K$, based on its brightness at 330 MHz and 
assuming it is optically thick.

  High resolution  interferometric RRL studies of various HII regions in
W51A have shown higher  electron  temperatures.  H109$\alpha$  studies by van
Gorkom et  al(1980)  found  $T_{e}$  for  G49.5-0.4  {\it  b,d,e}  sources to be
8500,6800,7400K  respectively;  H76$\alpha$  studies by Garay et al(1985)  found
6600K and 6400K for sources {\it d,e} respectively;  H96$\alpha$ observations
by Mehringer(1994)  found average values of $T_{e}$ around brightest  components
{\it d,e} in the  range of 7500K  and  6500K  respectively.  Similar  values  of
electron  temperatures have been reported from other HII regions by RRL studies;
e.g.  W3A:  7500 $\pm$ 750K( Roelfsema \& Goss,1991), SgrA West:  7000 $\pm$500K
(Roberts \&  Goss,1993).

  The discrepancy in the temperature estimates from the RRL and the low frequency flux measurements could be explained in different ways. If the HII region has fine structure or temperature gradients, the RRL and the radio continuum emission could arise from parts of the source with different properties. The low frequency continuum emission could be probing the outer parts of the HII region which could be cooler. If there is fine structure in the source, we can invoke the filling factor $f_{w}$ ranging from 1 to 3, to explain away the discrepancy. In Table 3, we have given the filling factor $f_{w}$ required to reconcile the temperatures obtained by two techniques. However, such structure should leave some observational footprint. We have examined the high resolution maps to see if we could find some correlation between the required filling factors and other properties of the source. The sources, G49.5-0.4{\it b,d,e}, that are strongest at 5 GHz, and are the ones with the highest emission measure, are also the ones requiring the smallest filling factor corrections. The source G49.5-0.4{\it h} for which we have a lower limit on the temperature is also the largest and most relaxed component where our measured brightness temperature is most likely to reflect the true brightness. Of the components with large filling factors, G49.5-0.4{\it a} shows fine structure, having a shell like structure. The component G49.5-0.4{\it c} is not very strong and is in a complex region and so could be a blend of different components. The components G49.5-0.4{\it (f,g)} are fairly relaxed and are comparable to our beam and there is no strong reason for the high filling factor, except that they are embedded in weaker diffuse emission.

	Our estimated values of the emission measure given in Table 3, are lower than those estimated from the RRL measurements. The fitting procedure gives a good estimate of the optical depth, but to convert it to emission measure one needs to know the temperature of the region. The low temperatures obtained by us using the raw fitting procedure leads to low values for the emission measure. We have recomputed the emission measures for the lowest temperature in the range given by the RRL measurements and presented these in column 5 ($EM'$) of Table 3. These are in general agreement with the RRL emission measures.

\begin{table} 
\begin{center} 
 
\caption{Derived Parameters ($T_{e},EM$) from Least-Square Fitting, RRL Temperature, Filling Factor, and Revised EM ($EM'$)}\vspace*{5mm} 
\begin{tabular}{cccccc}
\hline 
Source & $T_{e}$(K) & $EM$  & RRL $T_{e}$*(K) & $f_{w}$ & $EM'$\\
 
& & (10$^{6}$ cm$^{-6}$pc) & & & (10$^{6}$ cm$^{-6}$pc)\\ \hline
G49.5-0.4a & 2300$\pm$650 & 0.5$\pm$0.4 & 5500-7500 & $>2.4$ & $1.6$\\

b & 4300$\pm$600 & 3.8$\pm$1.8   & 5000-6500& $>1.2$ & $4.8$\\

c & 3200$\pm$550 & 1.2$\pm$0.6   & 6000-7500 & $>1.9$ & $2.8$\\

d & 4300$\pm$350 & 12.3$\pm$4.5  & 4500-7000 & $>1.1$ & $14.0$\\

e & 5600$\pm$450 & 15.6$\pm$4.8  & 6000-8000 & $>1.1$ & $17.7$\\

f & 2100$\pm$200 & 0.5$\pm$0.1   & & $>2.8$ & $2.0$\\

g & 2400$\pm$200 & 0.9$\pm$0.2   & & $>2.8$ & $3.6$\\ 

h & $\ge$4900 & $\ge$0.2   & & $>1.2$ & $0.3$\\
\hline
G49.4-0.3a & 2750$\pm$500& 0.5$\pm$0.3  & & $>2.2$ & $1.4$ \\

b & 3700$\pm$350 & 3.4$\pm$3.2  & & $>1.6$ & $6.4$\\

c & 3000$\pm$550 & 0.4$\pm$0.25  & & $>2.0$ & $1.0$\\ \hline
\end {tabular}

\flushleft\small{ * Temperatures quoted are from Mehringer(1994); for rest of sources, it is assumed that RRL temperature is 6000K.}

\label{tab3} \end{center} \end{table}

\begin{figure} 
     
\psfig{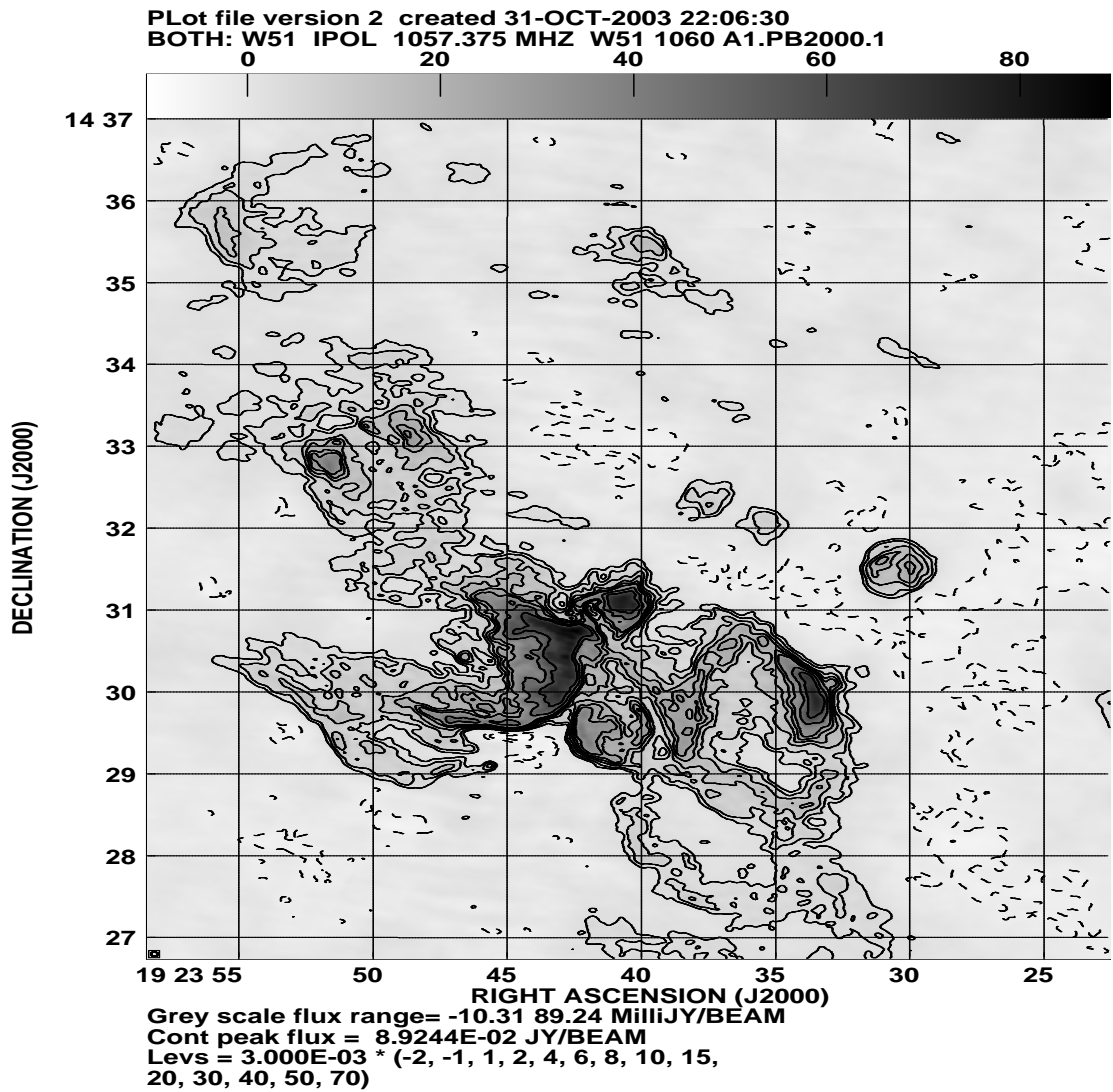}

\caption{  High resolution ($3.7''$$\times$$3.2''$ @ $82^\circ $) map of G49.5-0.4 at 1057MHz}
 
\end{figure}

\begin{figure}
  
\centerline{\hbox{
\psfig{figure=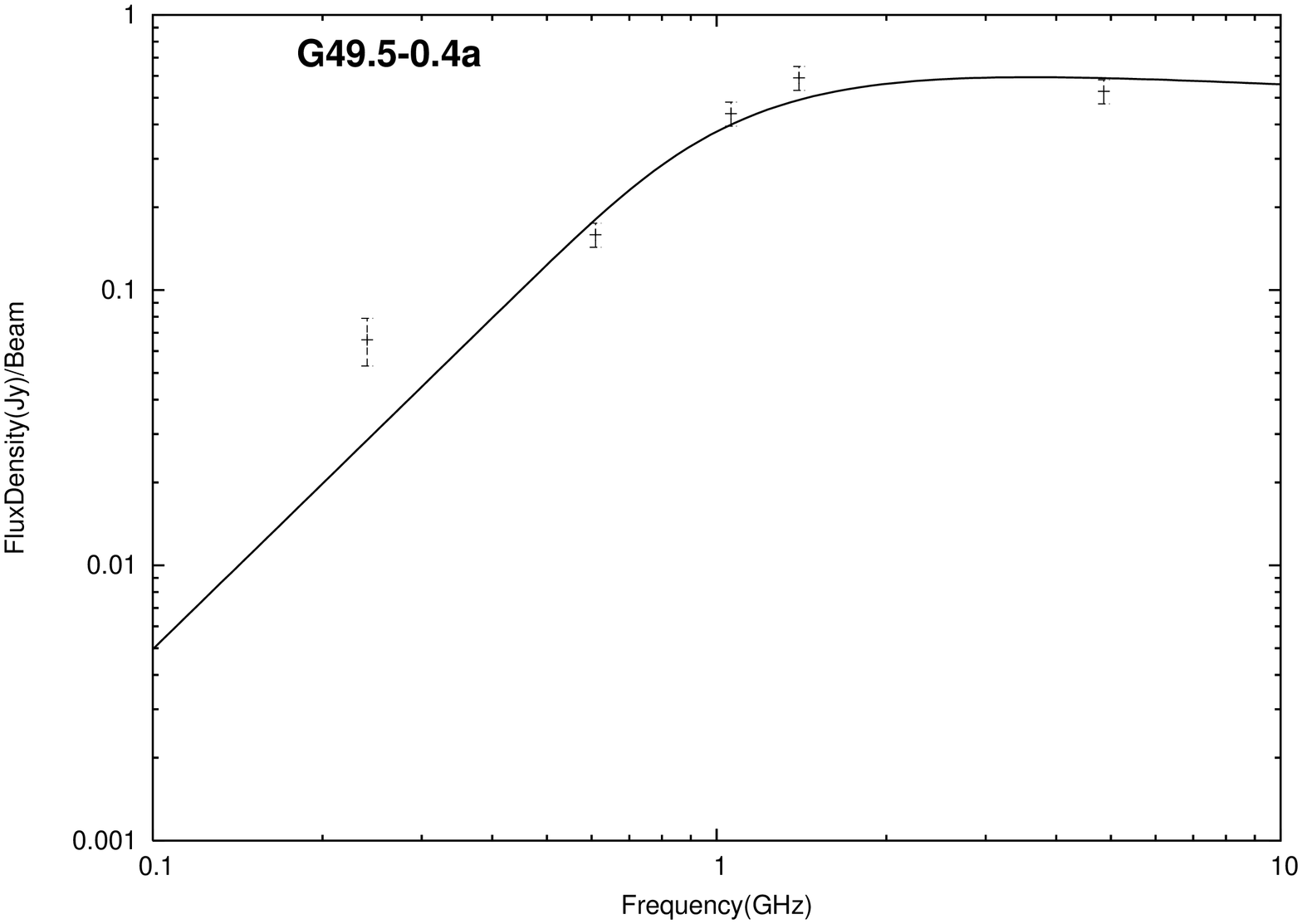,height=7cm,width=8cm,angle=-90}
\psfig{figure=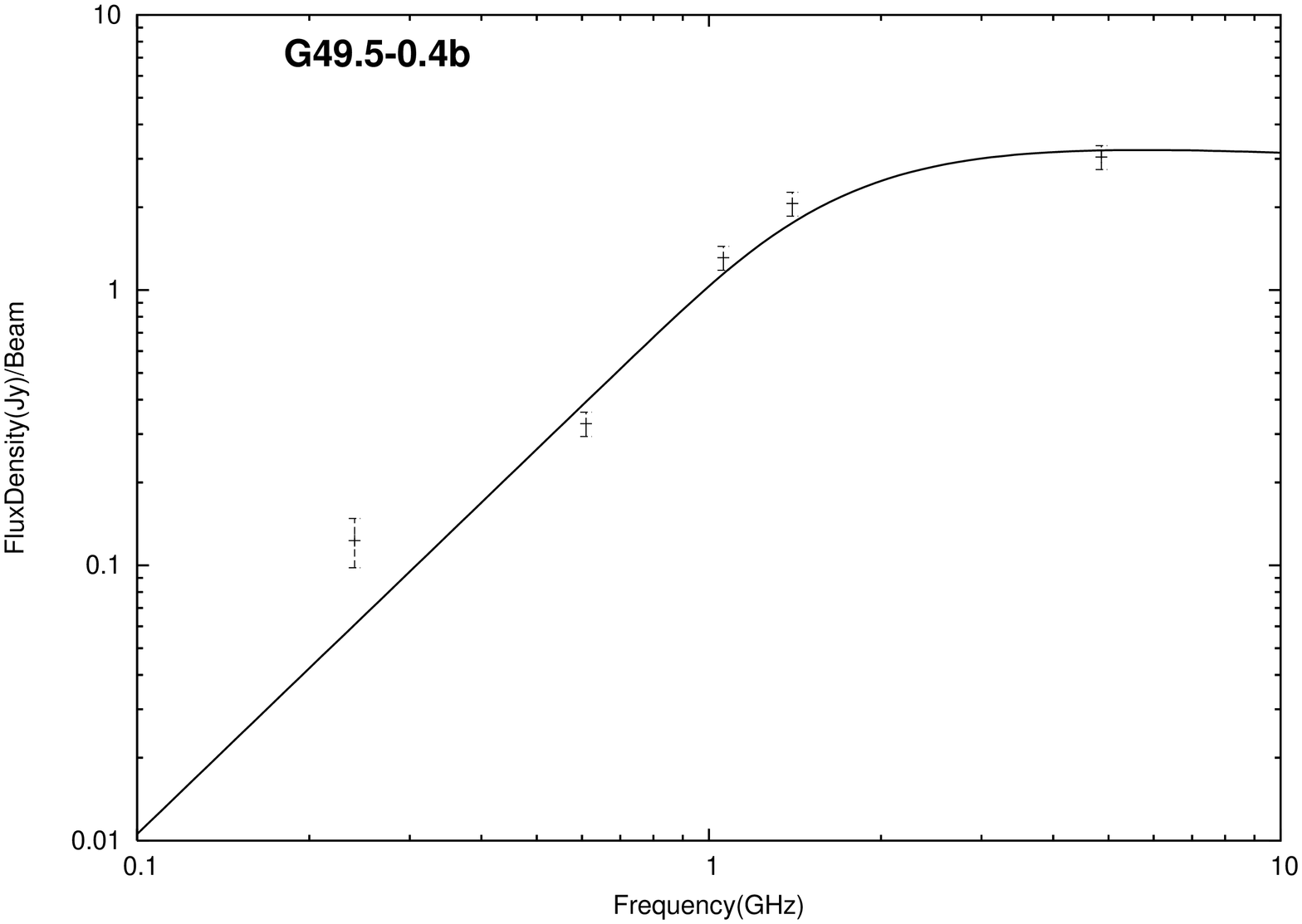,height=7cm,width=8cm,angle=-90}}} 
\centerline{\hbox{
\psfig{figure=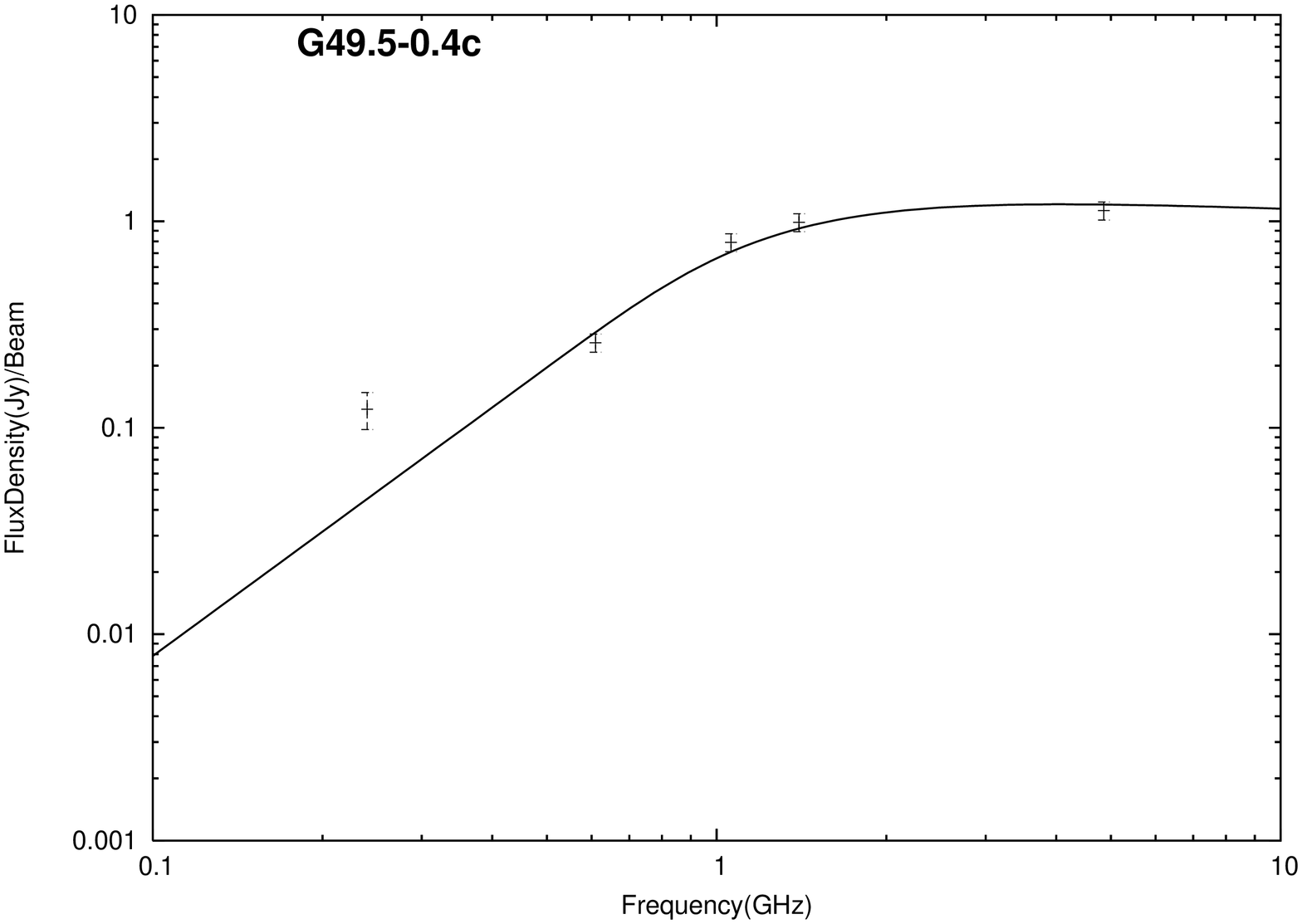,height=7cm,width=8cm,angle=-90}
\psfig{figure=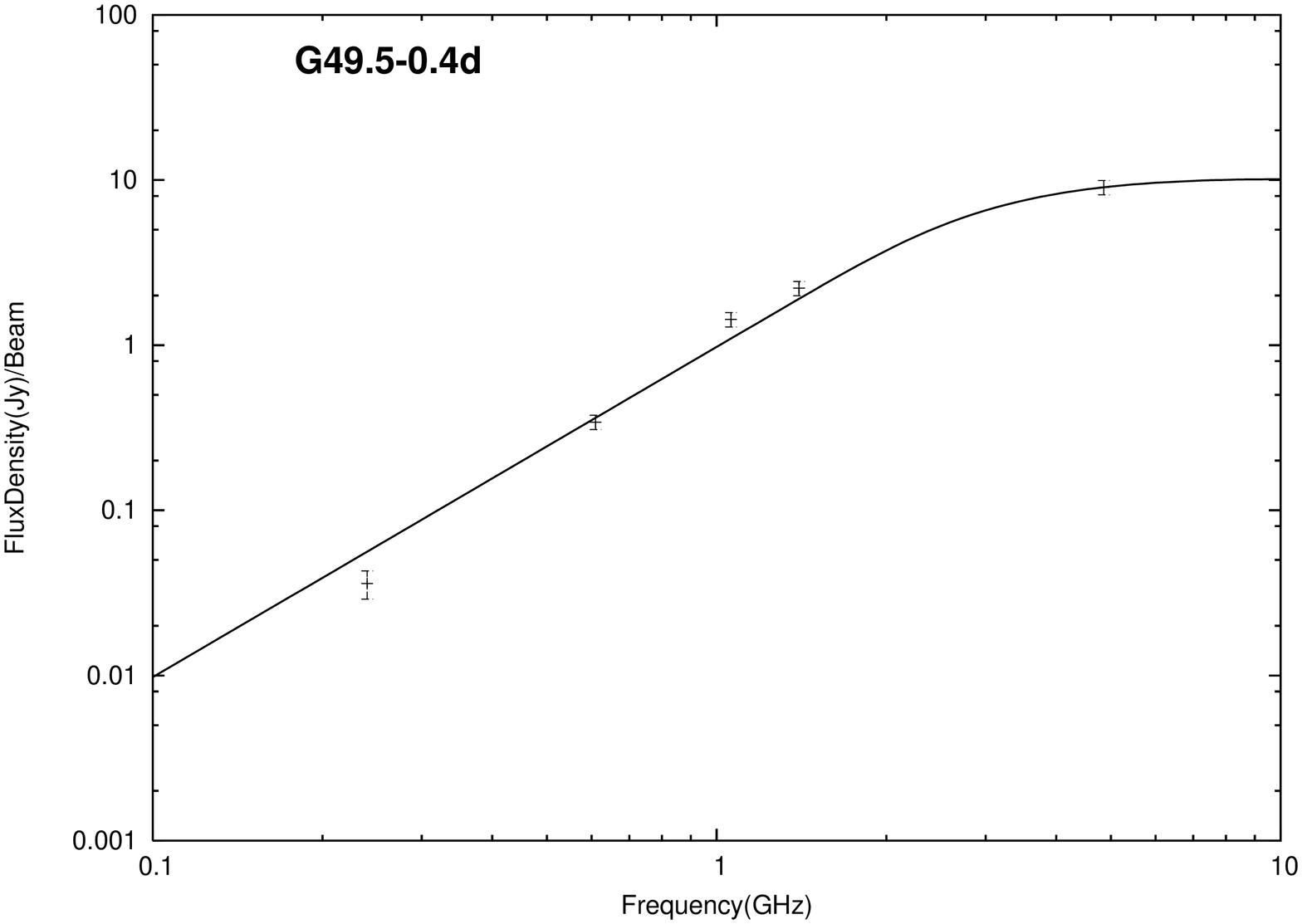,height=7cm,width=8cm,angle=-90}}} 
\centerline{\hbox{
\psfig{figure=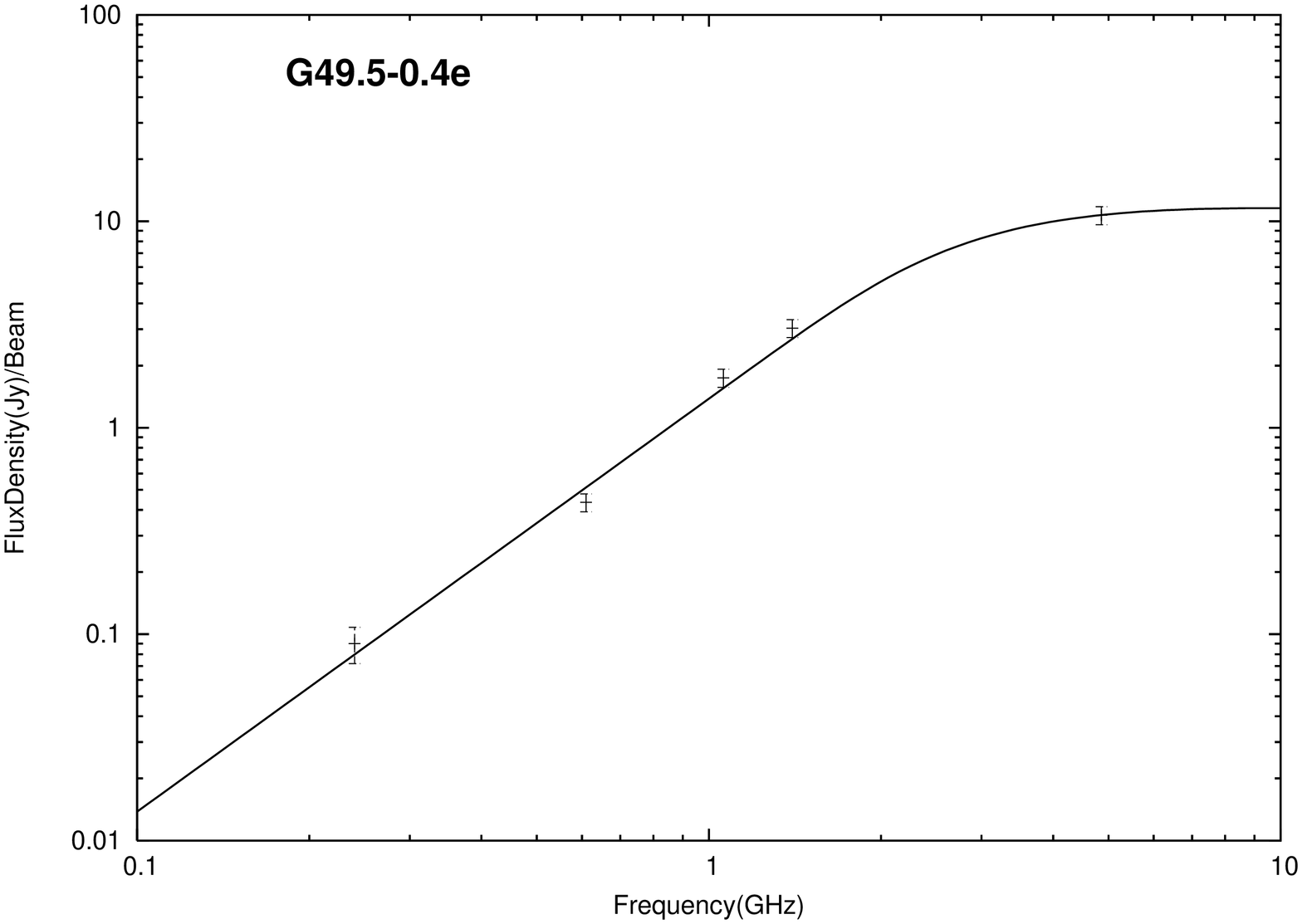,height=7cm,width=8cm,angle=-90}
\psfig{figure=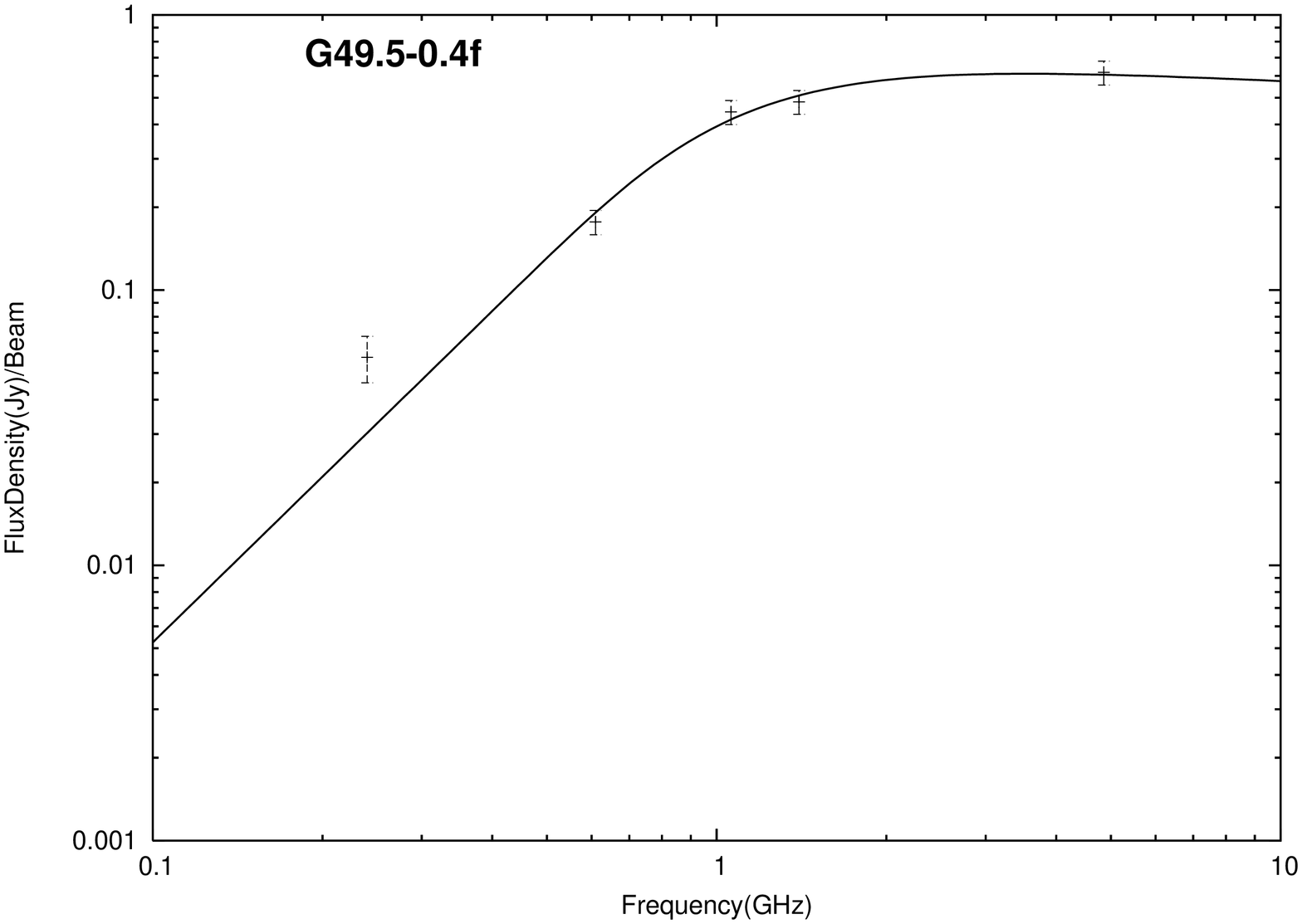,height=7cm,width=8cm,angle=-90}}} 
\end{figure}
\begin{figure}
\centerline{\hbox{
\psfig{figure=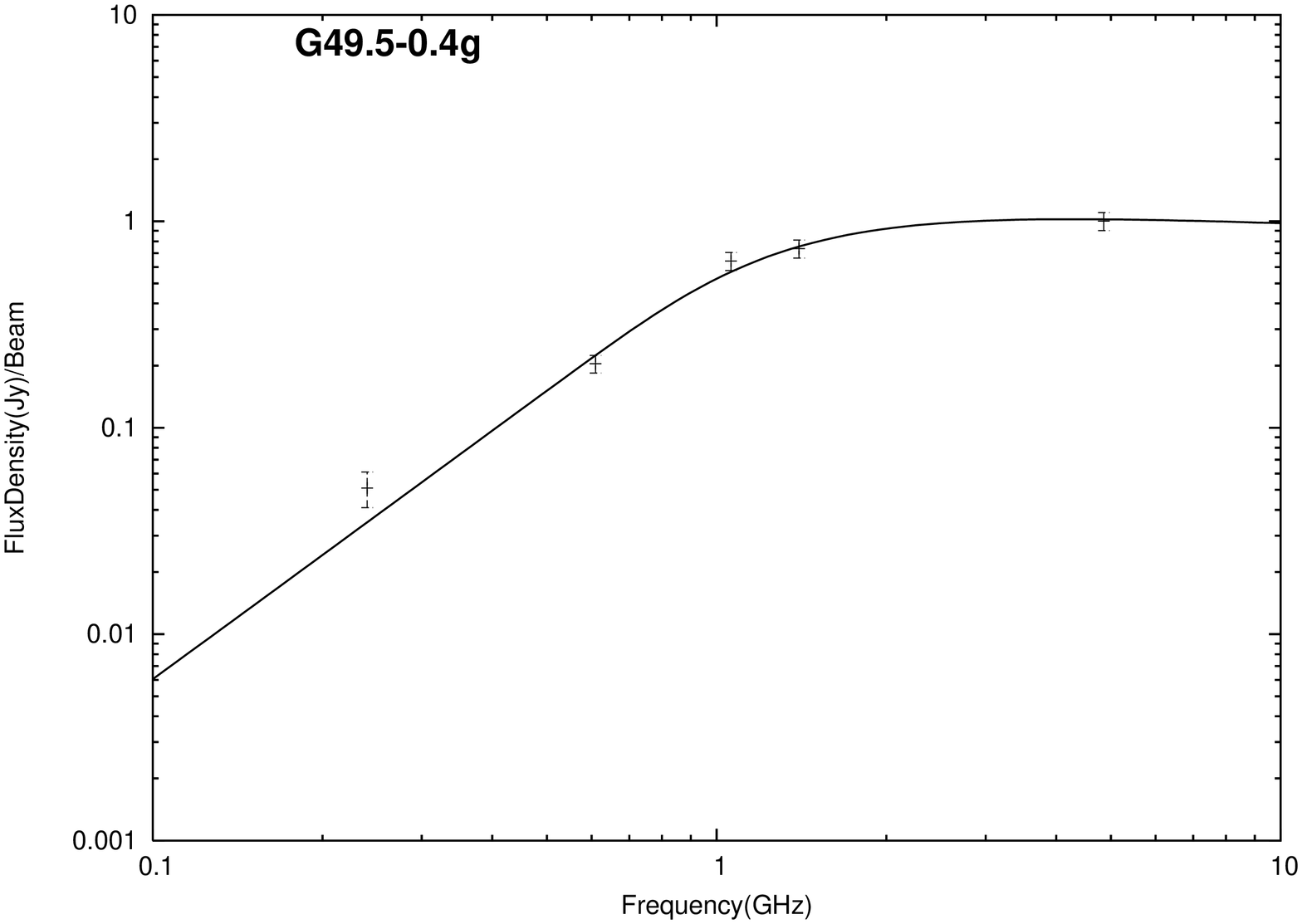,height=7cm,width=8cm,angle=-90}
\psfig{figure=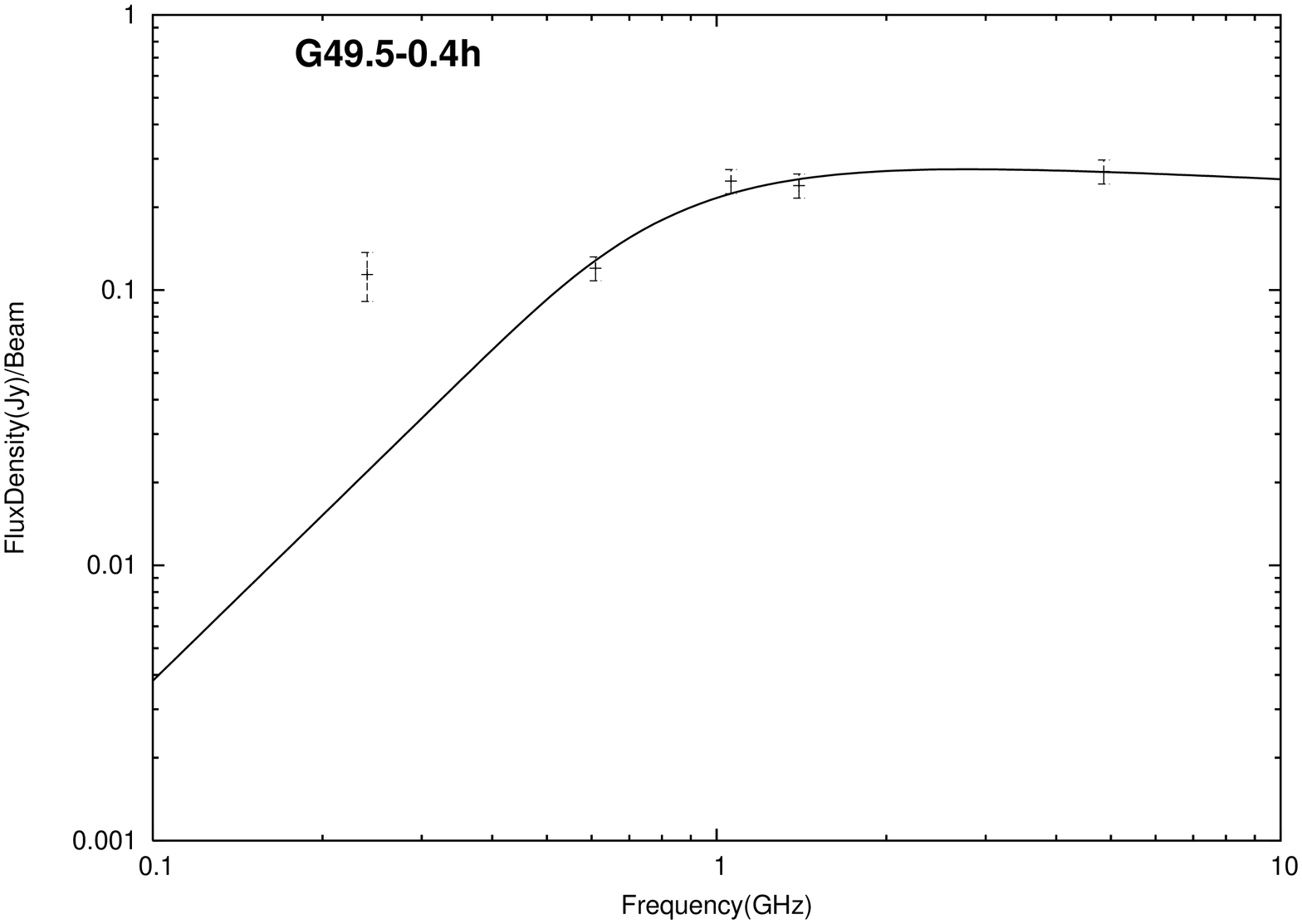,height=7cm,width=8cm,angle=-90}}} 

\caption {  Observed and fitted thermal  spectra  of HII regions in G49.5-0.4. Source {\it h} seems to be optically thin in the entire 
range}

\end{figure}

\begin{figure}  
\psfig{file=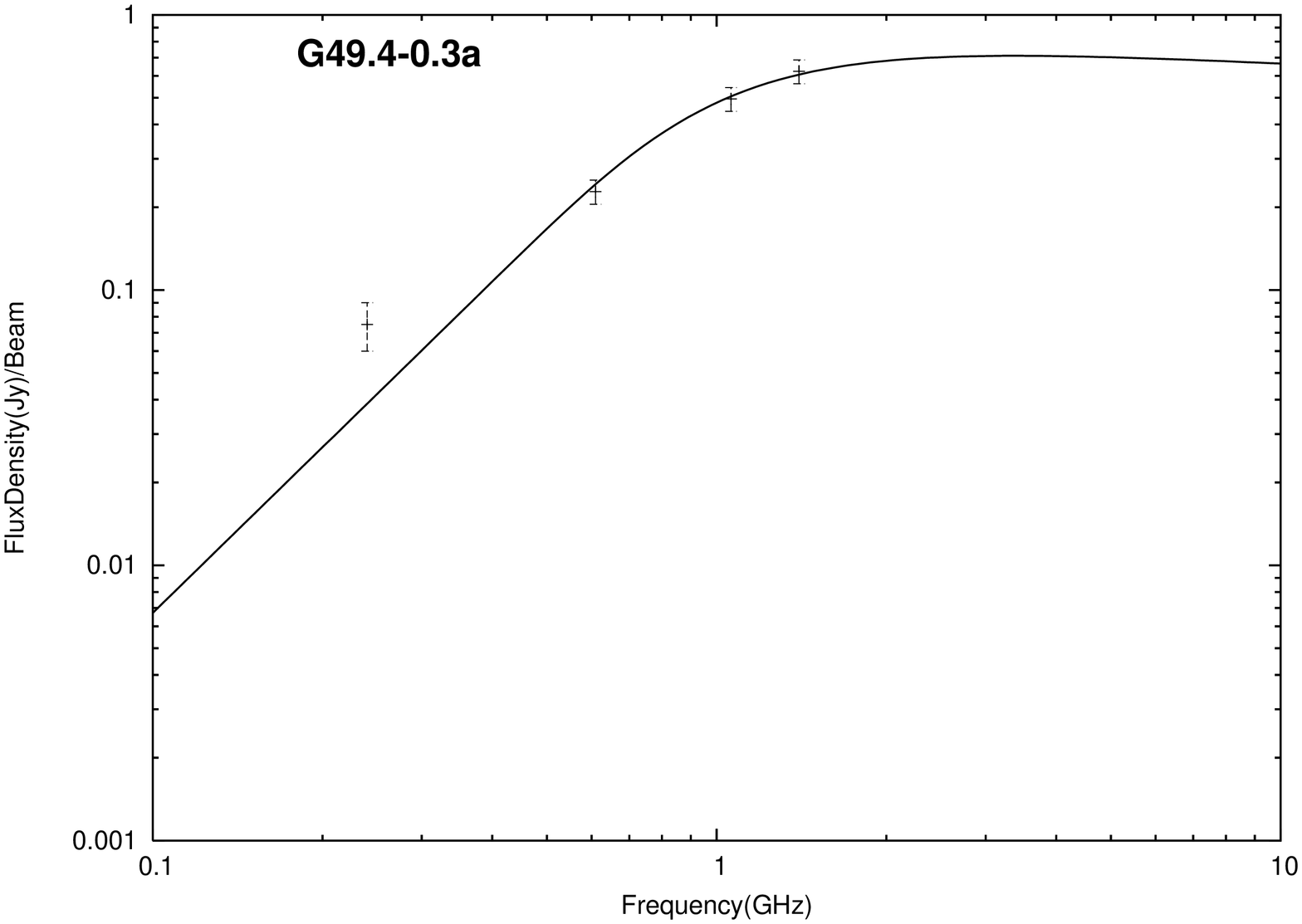,height=7cm,width=8cm,angle=-90}
\psfig{file=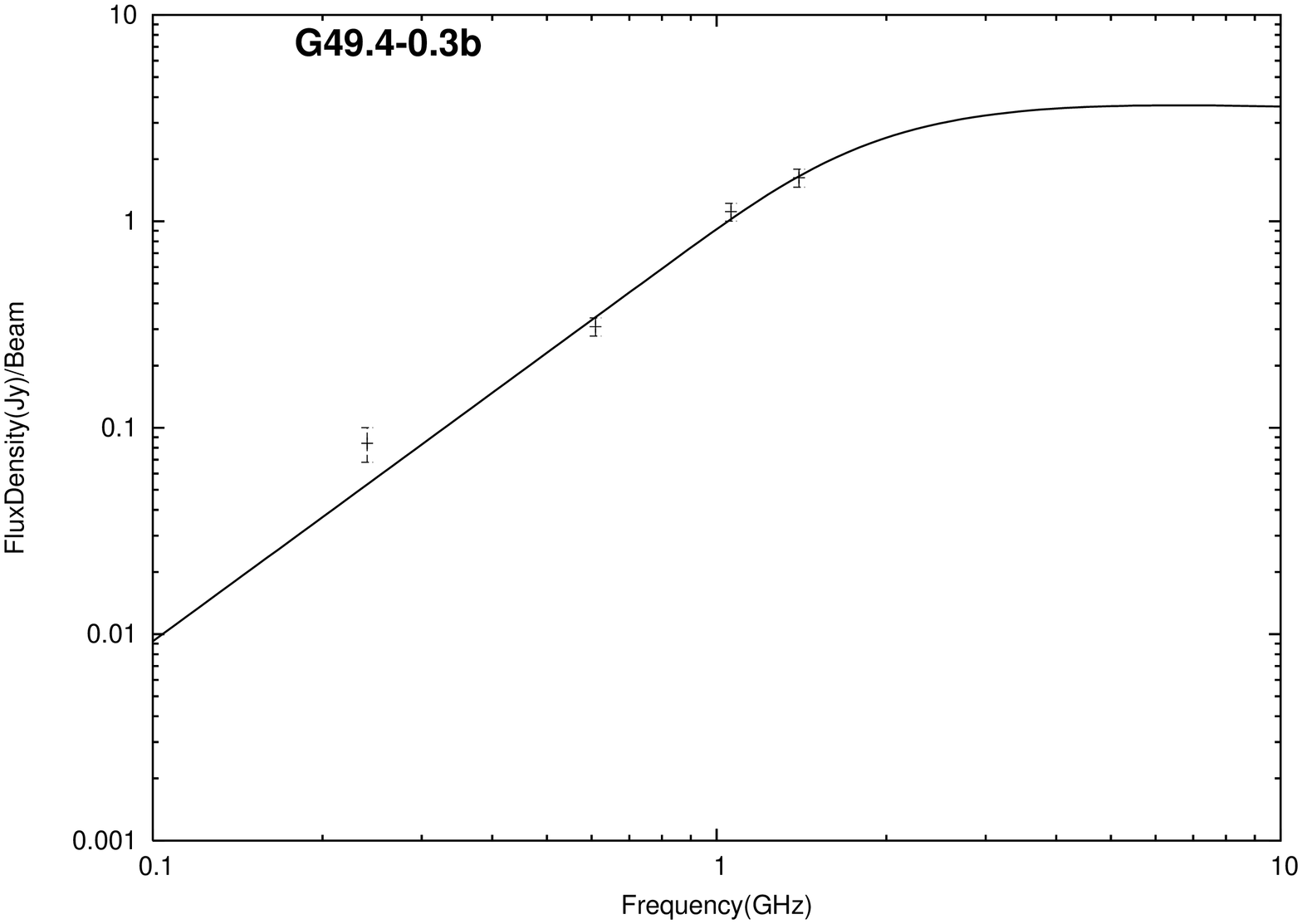,height=7cm,width=8cm,angle=-90} 
\psfig{file=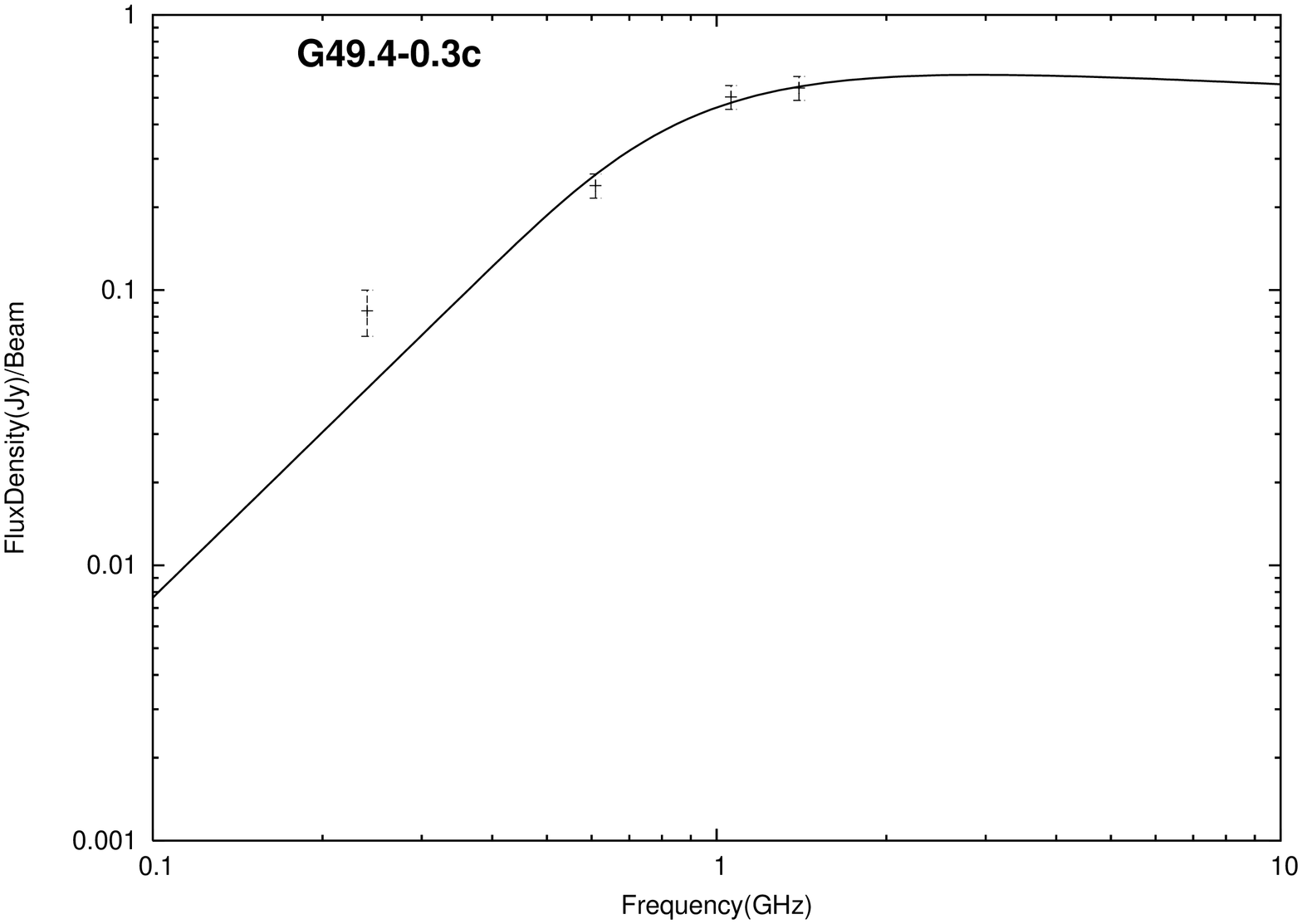,height=7cm,width=8cm,angle=-90}
\caption {  Observed and fitted thermal spectra  of HII  regions  in G49.4-0.3.}
\end{figure}

%\begin{figure}  
%\psfig{file=Fig4a.ps,height=10cm,width=10cm,angle=-90}\label {Fig.6}\
%\caption {  Thermal spectrum of HII  compact  source  in G49.5-0.4a with two-component structure. Solid line shows the spectrum of the  core as in Fig.4; dashed  line shows a  possible  spectrum  of the  outer hot envelope with Te=6200K and lower EM=0.2$\times$ 10$^{6}$cm$^{-6}$pc; dotted line shows the  resultant  which  accounts for  higher flux at 240MHz.}
%\end{figure}

\subsection {Excess Flux at 240 MHz and a Two Components Model}

	In many of the components like G49.5-0.4{\it a,b,c,f,g}
and all the components of G49.4-0.3, the brightness is higher at 240MHz than 
expected from the model fits.
The possibility of flux calibration errors at 240MHz was
ruled out by the general agreement of our flux density estimates of the components
of W51 and 4C13.73 with the expected values based on other observations
and the fact that the  
excess is not seen in some of the components like G49.5-0.4{\it d,e}. The physical location
of the GMRT antennas being the same for all the frequencies, the 240MHz observations have 
better short spacing coverage than at other frequencies and so will more faithfully reproduce the extended
emission. To avoid bias due to the extended emission, we have not directly used the observed peak brightness at 240MHz for the sources  but have made 1-dimensional cuts in different position angles and estimated the excess of the peak brightness over the local 
background. While this is a bit subjective, we believe that this should reduce the possibility
of systematically overestimating the brightness because of the the better short spacing coverage. 

For the components with this excess flux, if we consider only the 240MHz observations, the measured brightness temperature is comparable to the RRL temperatures. However, we do not think that this a resolution of the problem of the discrepancy between the RRL and the radio continuum temperatures since this does not take into account the observations at the higher frequencies. For complex structures, it is reasonable to expect the filling factor to depend on frequency, since different parts of the source get optically thick at different frequencies. One possible explanation is to consider a spherically symmetric model with a temperature gradient, but that would require the outer regions to be hotter than the inner which may not be reasonable, since the energy source is at the centre. Another possibility is a raisin pudding model, with dense regions embedded in hot plasma. While this may not be unreasonable, the non detection of such fine structure in the high frequency, high resolution maps (Mehringer 1994) is a concern.

Rather than associating the excess flux at 240 MHz to the compact sources, one could attribute it to foreground emission that becomes strong at low frequencies. Synchrotron emission, either from the galactic foreground or from relativistic particles in the vicinity, has the property of being stronger at lower frequencies; but any smooth large scale emission of this type will not be seen by interferometric observations. If interferometric observations picks up fine structure in this emission, then a chance positional coincidence with the foreground enhancement would be required to explain the observed 240 MHz excess. 

A more likely explanation is that the excess emission is due to an envelope of  hot low density thermal plasma associated with the HII complex, in which all the components are embedded. This component is optically thin at all the frequencies, including 240MHz and has a flat spectrum characteristic of optically thin thermal emission. This emission adds brightness to all frequencies, but at 240MHz the  emission of the optically thick inner components have fallen to low enough values for this foreground emission to be significant. For this explanation to work, the surrounding gas has to have a higher temperature than the compact components and should have structure on sufficiently small scale so that it is seen by the synthesis observations and and is not eliminated by our procedure of determining the peak brightness by taking its excess over the local mean. In the optically thin regime, we are not in a position to make independent estimates of the temperature and density of this component.

\section{Summary} 

	We have presented high resolution radio continuum images of the W51
complex made with the GMRT at 240, 610,1060 and 1400 MHz. While the 240MHz
observations cover the whole complex, the higher frequency observations 
have focussed on the thermal component known as W51A. All the subcomponents of
W51A are resolved in the $20''\times 15''$ images made at all the above frequencies. Combined
with VLA observations at 5GHz with comparable resolution, the peak brightness 
of all the components show a well defined thermal spectrum which is optically thick at frequencies less than 1 GHz. The emission
measures and the electron temperatures have been estimated by fitting the observed spectra to standard equation (Eq.3) of radiative transfer in HII regions. While the
estimated values for some of the components are in reasonable agreement with the estimates
from RRL observation, in many of the components, the continuum temperatures
are lower than the RRL values. This suggests that even with this resolution, there are unresolved features in these components and we have estimated the filling factors required to reconcile the two measurements.
The observed brightness of some of the components at 240MHz is higher than 
expected from the temperature predicted by the fit.
This could be due to a high temperature,
low emission measure envelope around the components which could be associated with the W51A complex. The discrepancies between the temperatures estimated from the continuum and the RRL observations and between continuum observations themselves suggest the need for more complex models for the HII regions including the 3-dimensional distribution of matter in the complex.

\section{Acknowledgements}

PKS acknowledges the support given by IUCAA under its Visiting Associate 
programme and by NCRA in providing its facilities.PKS also acknowledges the 
support given by University Grants Commission, India under Project 
F.8.4(74)/1999--2000/MRP.

\end{document}